\newif\ifrevision
\revisionfalse

\ifrevision

\newcommand{\added}[1]{\textcolor[rgb]{0.8,0,0}{#1}}

\else
\newcommand{\added}[1]{#1}

\fi

\documentclass[sigconf]{acmart}
\AtBeginDocument{%
  }

\usepackage{multirow}
\usepackage{tablefootnote}
\usepackage{listings}
\usepackage{enumitem}

\newcommand{\System}[0]{\textit{GestureCoach}}
\newcommand{\EmphModel}[0]{\emph{emphasis proposal}}
\newcommand{\GesModel}[0]{\emph{gesture identification}}
\newcommand{\numusers}[0]{10}

\acmYear{2025}
\setcopyright{cc}
\setcctype{by-nc}
\acmConference[UIST '25]{The 38th Annual ACM Symposium on User InterfaceSoftware and Technology}{September 28-October 1, 2025}{Busan, Republic of
Korea}
\acmBooktitle{The 38th Annual ACM Symposium on User Interface Software and
Technology (UIST '25), September 28-October 1, 2025, Busan, Republic of
Korea}
\acmDOI{10.1145/3746059.3747705}
\acmISBN{979-8-4007-2037-6/2025/09}

\begin{document}

\title{\System{}: Rehearsing for Engaging Talks with LLM-Driven Gesture Recommendations}

\author{Ashwin Ram}
\affiliation{%
  \institution{Saarland University,\\
Saarland Informatics Campus}
  \city{Saarbrücken}
  \country{Germany}
}
\email{ram@cs.uni-saarland.de}

\author{Varsha Suresh}
\affiliation{%
  \institution{Saarland University, \\
  Saarland Informatics Campus}
  \city{Saarbrücken}
  \country{Germany}
}
\email{vsuresh@lst.uni-saarland.de}

\author{Artin Saberpour Abadian}
\affiliation{%
  \institution{Saarland University,\\
Saarland Informatics Campus}
  \city{Saarbrücken}
  \country{Germany}
}
\email{saberpour@cs.uni-saarland.de}

\author{Vera Demberg}
\affiliation{%
  \institution{Saarland University,\\
Saarland Informatics Campus}
  \city{Saarbrücken}
  \country{Germany}
}
\email{vera@lst.uni-saarland.de}

\author{Jürgen Steimle}
\affiliation{%
  \institution{Saarland University,\\
Saarland Informatics Campus}
  \city{Saarbrücken}
  \country{Germany}
}
\email{steimle@cs.uni-saarland.de}

\renewcommand{\shortauthors}{Ram et al.}

\begin{abstract}

This paper introduces \emph{GestureCoach}, a system designed to help speakers deliver more engaging talks by guiding them to gesture effectively during rehearsal. \emph{GestureCoach} combines an LLM-driven gesture recommendation model with a rehearsal interface that proactively cues speakers to gesture appropriately. Trained on experts’ gesturing patterns from TED talks, the model consists of two modules: an emphasis proposal module, which predicts \emph{when} to gesture by identifying gesture-worthy text segments in the presenter notes, and a gesture identification module, which determines \emph{what} gesture to use by retrieving semantically appropriate gestures from a curated gesture database. Results of a model performance evaluation and user study (N=30) show that the emphasis proposal module outperforms off-the-shelf LLMs in identifying suitable gesture regions, and that participants rated the majority of these predicted regions and their corresponding gestures as highly appropriate. A subsequent user study (N=10) showed that rehearsing with \emph{GestureCoach} encouraged speakers to gesture and significantly increased gesture diversity, resulting in more engaging talks. We conclude with design implications for future AI-driven rehearsal systems.

\end{abstract}
%
%


%

\begin{teaserfigure}
\centering
  \includegraphics[width=\textwidth]{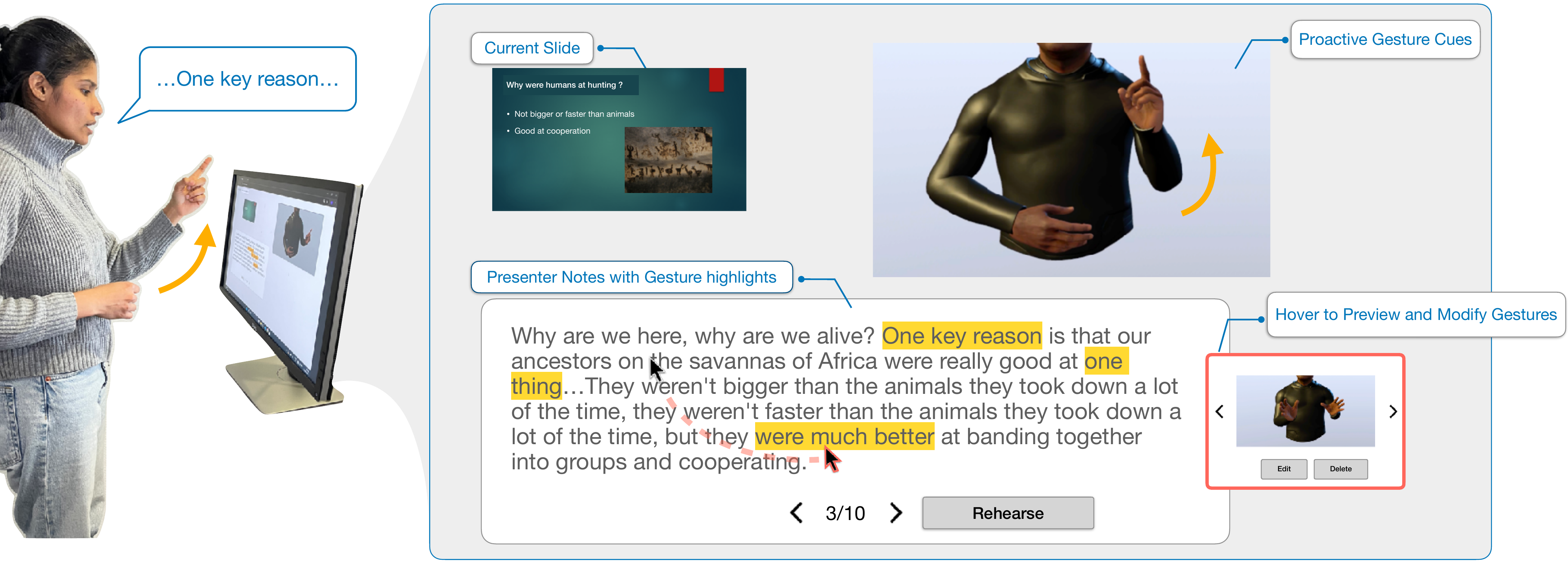}
  \caption{\System{} guides speakers to perform gestures while rehearsing their talk. A gesture recommendation model predicts text segments in the presenter notes that should be emphasized with gestures and retrieves relevant semantic gestures for each segment. During rehearsal, the system highlights the segments and proactively cues a video clip of the gesture by tracking users' speech, allowing them to integrate the gesture smoothly into their talk. Hovering over a segment allows users to preview and modify the associated gesture with alternate suggestions from the model.}
  \Description{The left side of the figure shows a speaker rehearsing for a presentation talk with the GestureCoach system. The speaker is saying "one key reason" and showing the gesture symbolizing one by following a gesture cue provided by the system where a humanoid avatar demonstrates the same gesture. The right side shows a zoomed in view of the rehearsal interface. The interface has the appeal of the presenter view mode typically found in presentation software. It has 4 main windows. 1) the current slide 2) the presenter notes for the slide with text segments where the speaker should perform gestures highlighted in yellow. This window also has a rehearse button that speakers click to start rehearsal. 3) a window where a humanoid avatar displays the gesture that should be performed for each highlighted region. 4) a popup window which appears when speakers hover over a highlighted region. This window shows video clips of the top-3 gesture recommendations provided by the recommendation model as a carousel design. It has an edit and delete button that allows users to browse and select the gesture they prefer for a region or remove the region if they do not prefer to gesture there.}
  \label{fig:teaser}
\end{teaserfigure}


\maketitle

\section{Introduction}

Delivering a “great” talk requires more than just good content; it needs an engaging delivery with the right blend of non-verbal cues, such as gestures~\cite{goldin2013gesture,krauss1996nonverbal,mcneill2008gesture}. In particular, semantic gestures that complement the spoken content \cite{mcneill1992hand}, such as tracing the outline of a circle in the air while saying 'a round table' or moving hands upward while saying 'rising prices', play a critical role in enhancing the dynamism of the speech ~\cite{burgoon1990nonverbal,rodero2022effectiveness,kendon1994gestures} and facilitating comprehension ~\cite{maricchiolo2020effects,kendon1994gestures,hostetter2011gestures}.

However, speakers, especially novices, often struggle with using semantic gestures in their talks \cite{so2013speakers,zeng2022gesturelens}. They may either completely neglect to use gestures or rely heavily on beat gestures ~\cite{mcneill1992hand}—simple, repetitive hand movements that do not enhance the meaning of the content. While some existing training systems ~\cite{schneider2015presentation, damian2015social, barmaki2016kinesic, tanveer2015rhema} track users’ motion and provide generic feedback to use more or fewer gestures, they do not offer specific guidance on what semantic gesture to use or when to use it for effective emphasis. Prior research has shown that verbal qualities, such as voice modulation, can be improved with rehearsal and then effectively applied during live talks ~\cite{wang2020voice}. Inspired by this work, we aim to develop a rehearsal system specifically designed to improve semantic gesture usage during talks.

Designing a rehearsal system that can give targeted guidance for semantic gestures, however, presents various technical and design challenges. The key difficulty lies in developing a gesture recommendation model that predicts appropriate text segments in the presenter notes that should be emphasized with gestures (henceforth referred to as gesture regions) and retrieves suitable semantic gestures for effective emphasis. Predicting whether a text segment is a gesture region depends on several factors, including its broader context within the text, the topic of the talk, and implicit knowledge expert speakers use to engage an audience. Furthermore, the suggested gestures must extend beyond the limited gesture vocabulary of novices, offering access to the more refined and expressive gestures used by experts. From a design perspective, these gesture recommendations must be delivered proactively during rehearsal, allowing users to practice and seamlessly incorporate them into their talk.

As a step towards tackling these challenges, we propose \System{}, the first system capable of providing targeted guidance to perform gestures during rehearsal. \System{} combines a gesture recommendation system with a rehearsal interface that delivers real-time gesture cues, helping speakers naturally integrate gestures while speaking. The gesture recommendation system utilizes recent advances in AI chaining, by combining two specialized Large Language Model (LLM) modules, each targeting a specific aspect of gesture-based emphasis. First, the \emph{emphasis proposal} module takes presenter notes as input and determines appropriate gesture regions. For this, we leverage the foundational knowledge of LLMs and further fine-tune it with annotated data from 10 TED talk videos of expert speakers. Second, the \emph{gesture identification} module selects the specific semantic gesture to emphasize each predicted gesture region \added{in a data-driven manner,} retrieving relevant examples from a curated database of gestures performed by experts using a Retrieval-Augmented Generation (RAG) framework ~\cite{lewis2020retrieval}.

We then identified key design requirements for a gesture rehearsal system through an iterative design process, which included developing and evaluating an early Wizard-of-Oz prototype in a pilot study with five participants. This study highlighted the need for an effective Human-AI collaborative interface design; one that enables users to proactively engage with and co-create the AI's gesture recommendations. Based on these findings, we developed a desktop-based rehearsal interface (shown in Figure ~\ref{fig:teaser}) that tracks users’ speech in real-time and proactively cues users during their practice as and when gestures need to be performed. Additionally, the system supports collaboration by allowing users to modify the AI's recommendations, enabling the creation of personalized gesture palettes tailored to each speaker’s preferences.

We evaluate our system in two ways: 1) technical evaluation of our gesture recommendation model, and 2) user evaluation of the final \System{} system. Results show that our emphasis proposal module outperforms off-the-shelf LLMs in identifying appropriate gesture regions in presenter notes. To assess the quality of these predictions, particularly false positives, we conducted a user study (N=30) evaluating both the appropriateness of the gesture regions and the suitability of the semantic gestures from the gesture identification module. Participants rated the majority of predictions as highly appropriate and the gestures as well-suited for conveying emphasis. Finally, in a user study (N=10), participants rehearsed for a talk using \System{}. Results show that speakers preferred using \System{} over rehearsing with just presenter notes, and they performed significantly more unique semantic gestures after using the system. Users also found the system easy to use and effective, reporting that it encouraged them to use a more diverse range of gestures and deliver more engaging talks. We also identified current limitations and outline opportunities for future work to improve AI-driven rehearsal systems and advance human-AI collaborative interfaces more broadly.

Our contributions include:
\begin{itemize}[leftmargin=*, noitemsep, topsep=3pt]
    \item We develop a gesture recommendation model and evaluate its performance for identifying gesture regions and for suggesting semantic gestures given input presenter notes.
    \item We implement an interactive rehearsal interface designed to help speakers integrate semantic gestures when giving a talk.
    \item Insights from a user study with \System{} on the strengths and limitations of AI-driven gesture rehearsal and design implications for future systems in this space.
\end{itemize}

\section{Related Work}

\begin{figure*}[th]%
\centering
\includegraphics[width=\linewidth]{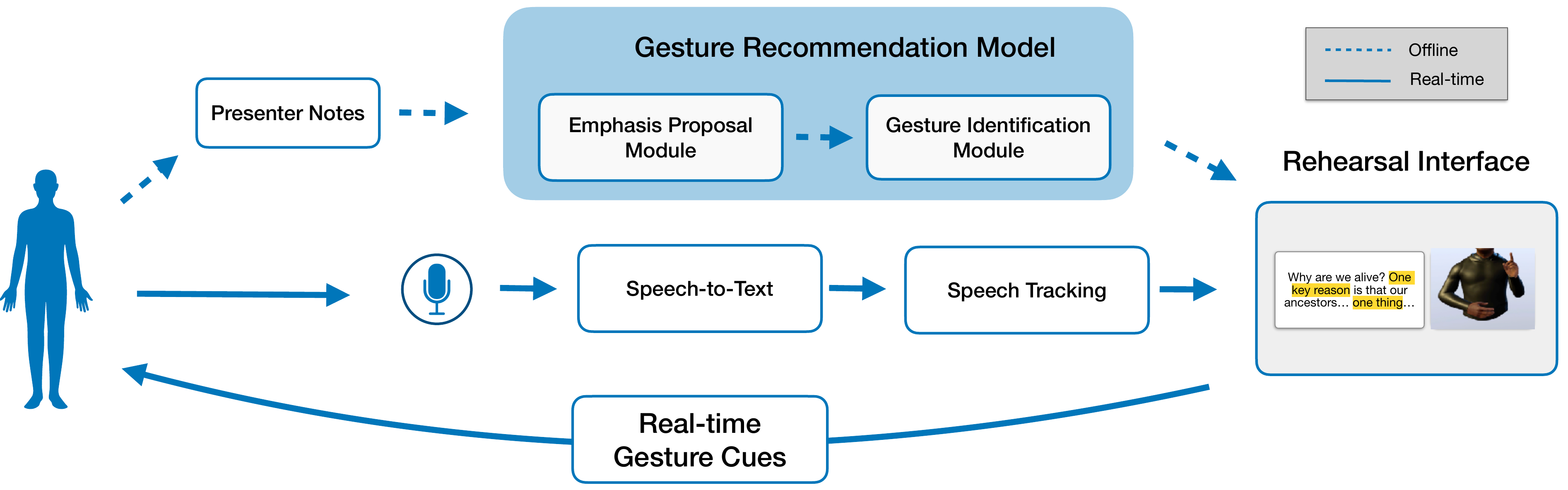}
\caption[]{Overview of the \System{} system. Users' presenter notes are processed by a gesture recommendation model in the backend. A frontend  rehearsal interface displays the recommended gesture regions and gestures. During rehearsal, users' speech is tracked and gesture cues are proactively delivered to provide real-time gesture guidance. }
\Description{The figure shows a diagrammatic overview of the various components in the GestureCoach system and how they connect together. The presenter notes provided by the user is analysed by the Gesture recommendation model and the predicted gesture regions and gesture cues are passed to the rehearsal interface. During rehearsal, the interface tracks users' speech to monitor their progress in the presenter notes and real-time gesture cues are fed back to user so that they can integrate them in their talk.}
\label{fig:system-overview}
\end{figure*}

\subsection{Co-speech Gestures for Communication}
\label{sec:gestures_overview}
Co-speech gestures are hand and arm movements that naturally accompany speech and play a vital role in communication \cite{kendon1972some, mcneill1992hand, shattuck2019dimensionalizing, maricchiolo2020effects}. These gestures are broadly classified into four types: beat, iconic, metaphoric, and deictic gestures \cite{mcneill1992hand}. Beat gestures are rhythmic movements that align with speech prosody to emphasize its structure. In contrast, iconic, metaphoric, and deictic gestures are semantic gestures that are directly related to the linguistic content being spoken, \added{enhancing content delivery during communication ~\cite{maricchiolo2020effects,kendon1994gestures,hostetter2011gestures}.}  Iconic gestures physically resemble the objects or actions they represent, such as mimicking an object's shape or reenacting a movement. Metaphoric gestures convey abstract concepts through spatial metaphors, like cupped hands symbolizing "holding" an idea. Deictic gestures involve pointing to indicate referents, whether physically, such as gesturing toward an object, or abstractly, such as indicating an imaginative space.

Semantic gestures, which are the focus of this study, \added{have been empirically shown to significantly benefit listener comprehension, especially when they depict motor actions or convey non-redundant information \cite{hostetter2011gestures}.} These gestures also help speakers express their intentions through complementary movements, making the message clearer and more engaging for listeners \cite{mcneill1992hand, kendon2004gesture, hinnell2019verbal}. For example, moving one’s hands from side to side can indicate contrast when saying “on the other hand” \cite{hinnell2019verbal}. Moreover, encouraging speakers to use hand gestures has been found to increase narrative length and influence prosodic features, such as fundamental frequency and intensity, thereby improving intonation \cite{cravotta2019effects}. These findings show that semantic gestures play an important role in making communication and presentations more effective.

\subsection{Systems for Rehearsing Talks}

Tutoring systems aimed at improving task-related skills span various domains, including machine task training in manufacturing \cite{huang2021adaptutar,liu2023instrumentar,stanescu2022model}, surgical training  \cite{faridan2023chameleoncontrol,ferrier2023learning}, and sports training through real-time feedback \cite{ma2024avattar,liu2022posecoach,geisen2022real}. In the domain of public speaking, such as giving talks or presentations, training systems have predominantly focused on vocal delivery, offering feedback on aspects like intonation and speaking pace \cite{tanveer2015rhema, wang2020voice, damian2015social, trinh2017robo}. Despite the well-established role of gestures in enhancing communicative effectiveness, there has been relatively little research dedicated to systems that support the development of gesturing skills while giving talks.
Although a few talk rehearsal systems also assist presenters in improving their gestures, they typically provide high-level guidance rather than detailed recommendations. For example, some approaches monitor a speaker’s body posture to detect closed positions or limited gestural variation, and provide suggestions to gesture more or less \cite{schneider2015presentation, damian2015social, barmaki2016kinesic}. While this in-situ feedback encourages speakers to incorporate or reduce gestures, it does not provide specific guidance on \textit{when} to gesture or \textit{what} gestures to use for effective emphasis.

Recent advancements in generative AI, particularly LLMs, have opened up new possibilities for building more intelligent interactive systems. In HCI research, the versatility of LLM-based tools has been demonstrated across diverse domains such as writing assistance \cite{zhang2023visar, xu2024skip}, image generation \cite{brade2023promptify}, programming \cite{kaze2023code, angert2023spell}, cognitive support \cite{zulf2024memoro}, choreography \cite{liu2024dance}, content creation in AR/VR \cite{su2024sonify, tan2024audio, torre2024llm}, and natural language-driven physical design \cite{qian2024shape}. More recently, LLMs have been applied to presentation training; for example, Park et al. introduced a system that generates diverse virtual audiences to provide feedback on talks \cite{park2023audilens}.

Building on this foundation, our work introduces an LLM-driven rehearsal system that offers fine-grained, proactive guidance for integrating meaningful semantic gestures. By analyzing presenter notes, the system identifies points where gestures would be most impactful and recommends specific gestures based on patterns drawn from expert speakers. This approach provides more targeted support than existing systems, enhancing speakers’ ability to deliver expressive and engaging presentations.

\subsection{Modeling Semantic Gestures}

Modeling semantic gestures in co-speech gesture synthesis is challenging due to the complex interaction between language and human motion \cite{yoon2022genea,nyatsanga2023comprehensive}. Unlike beat gestures, semantic gestures are sparse and exhibit a long-tailed distribution, making it challenging for models to learn their patterns effectively \cite{yoon2019robots,kucherenko2020gesticulator,mughal2024retrieving}. There has been active research on modeling semantic gestures for co-speech gesture synthesis \cite{zhang2024semantic,zhi2023livelyspeaker,mughal2024convo,mughal2024retrieving,liang2022seeg,kucherenko2020gesticulator,cheng2024siggesture}, with most efforts focusing on generating gestures from speech. 

To specifically incorporate semantic gestures into the gesture generation process, some works predict gesture properties such as gesture type information (deictic, metaphoric, and iconic) and gesture phase information from speech, in order to condition or retrieve semantic gestures for co-speech gesture generation \cite{kucherenko2020gesticulator, mughal2024retrieving}. However, when it comes to rehearsing a presentation where the user only has presenter notes and not the actual speech, applying these methods becomes challenging. A recent work predicts semantic gestures using labels such as "hands chop" and "throat cut" in location of the text transcript, using a predefined motion library \cite{zhang2024semantic}. However, the semantic gestures in this approach are limited to the library's set, which does not cover domain-relevant gestures, such as those performed while giving presentation talks. 

In this work, we aim to disentangle the tasks of predicting \textit{when} to gesture and \textit{what} semantic gesture to perform \cite{neff2016hand}. Linguistic research suggests that semantic gestures often co-occur with specific spoken discourse markers, such as discourse connectives (e.g., “however,” “for example”) that indicate topic shifts, and stance markers (e.g., “absolutely” “probably”) that reflect the speaker’s level of certainty \cite{mcneill2014discourse, laparle2021discourse, calbris2011elements,suresh2025enhancing}. For instance, raised index fingers tend to co-occur with contrastive discourse markers like "but," "however," that signal exceptions \cite{inbar2022raised}. These insights suggest that there are detectable patterns that can help determine when to gesture directly from text, which we address by fine-tuning an LLM to predict these gesture regions. Determining what semantic gesture to perform, \added{on the other hand, is more challenging as there are no clear rules on what gesture best fits a given context. Hence, we guide our modeling using a data-driven approach, dynamically retrieving contextually appropriate gestures from a corpus of expert gestures using RAG.}

\section{\System{}: Overview}

We present \System{}, a rehearsal system designed to help speakers practice their presentations while integrating semantic gestures effectively. Unlike previous training systems that offer generic cues to increase or decrease gesturing during talks ~\cite{schneider2015presentation, damian2015social, barmaki2016kinesic}, \System{} provides targeted guidance by analyzing expert speakers’ behavior to determine both \emph{when} a speaker should gesture during their talk and \emph{what} semantic gesture they should use for effective emphasis.

Figure ~\ref{fig:system-overview} provides an overview of our system. To start with, the speaker provides their presenter notes as input. \System{} then processes the presenter notes to identify specific text segments where gestures can enhance clarity and emphasis, i.e., gesture regions, and suggests appropriate semantic gestures for each region. As the speaker rehearses by reading the notes aloud, the interface actively tracks their speech to identify their position in the notes and proactively plays a video clip of the semantic gesture to be performed in real-time. In the following sections we elaborate on the two main components of \System{}: 1) a backend gesture recommendation model (Section ~\ref{system:model}), and 2) a frontend rehearsal interface (Section ~\ref{system:interface}).

\section{\System{}: Recommendation Model}
\label{system:model}

Developing an effective gesture recommendation model that predicts both \emph{when} a speaker should gesture and \emph{what} semantic gesture to use for emphasis presents a significant challenge. Gesture placement and selection are not merely a function of the words spoken; rather, they require a deeper knowledge of the contextual meaning, discourse structure, and audience engagement strategies that skilled speakers intuitively employ. 

Given recent advancements in language modeling, we hypothesized that zero-shot prompting of an off-the-shelf LLM such as GPT-4o could effectively identify gesture regions in presenter notes where speakers should gesture. However, our preliminary testing revealed two key challenges. First, the LLM predicted large segments of text as gesture regions, likely because it lacks knowledge of expert gesturing patterns. This lack of precision can make it difficult for users to determine exactly when to gesture while speaking. Second, it provided generic or abstract descriptions of what gestures to perform rather than specific, actionable suggestions. This lack of clarity, especially for novice speakers, can make it challenging to translate the recommendations into meaningful physical movements.

\begin{figure*}[t]%
\centering
\includegraphics[width=\linewidth]{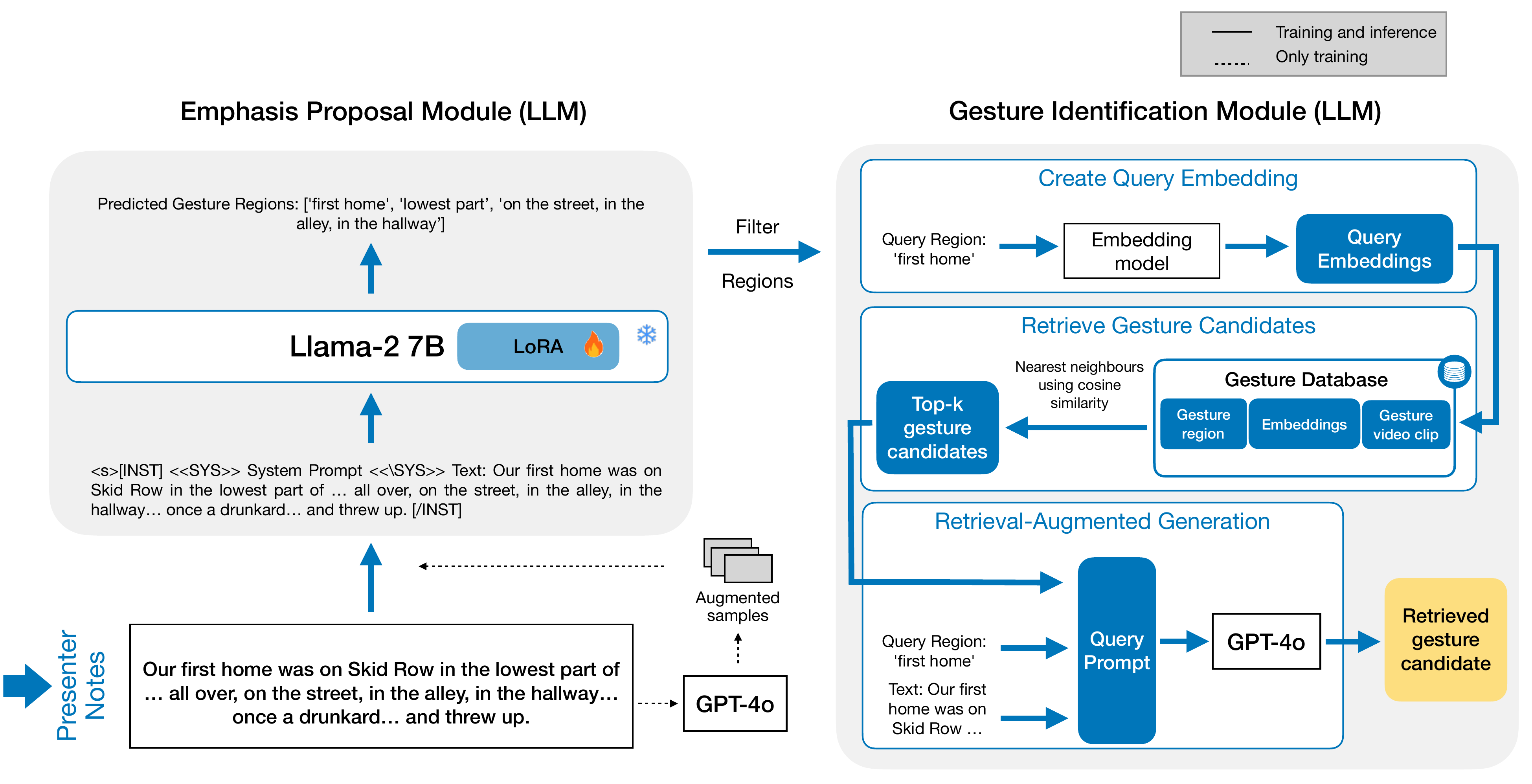}
\caption[]{Gesture recommendation model architecture. It consists of two LLM-based modules: 1) The \emph{emphasis proposal} module is fine-tuned on data from expert speakers to predict gesture regions in the presenter notes. The predicted regions are filtered and passed to 2) the \emph{gesture identification} module which selects the most suitable semantic gesture from the gesture database using Retrieval-Augmented Generation.}
\Description{A diagrammatic figure detailing the architecture of the gesture recommendation model. The left side shows the flow for the emphasis proposal module. A sample presenter notes is shown and is provided as input to this module. The module predicts gesture regions in the notes as output. These regions are filtered and passed to the gesture identification module shown on the right side. For each predicted gesture region, a vector embedding is created using an embedding model. The embedding is used to query a gesture database that contains data from expert gesturing patterns. The top-k most similar gesture candidates from the database is retrieved and passed to an LLM, along with the predicted gesture region and the notes to select the most suitable gesture for the region from the candidates.}
\label{fig:model}
\end{figure*}

To address these issues, we developed a gesture recommendation model that accurately predicts specific parts of the presenter notes that should be emphasized with gestures, and provides corresponding semantic gesture in the form of short (1-3 seconds) video clips. Figure ~\ref{fig:model} shows an overview of our model.  It comprises of 1) a gesture database created from TED talk videos, by annotating the text segments in the talk transcript where speakers performed semantic gestures in the video. 2) an \emph{Emphasis Proposal} module, trained on this database to identify gesture regions in the presenter notes that should be highlighted with semantic gestures. 3) a \emph{Gesture Identification} module that retrieves a suitable semantic gesture for each predicted gesture region from a database of recorded gesture videos.

\subsection{Gesture Database}
\label{sec:database}

To create a gesture database, we first collected TED talk videos of expert speakers who performed semantic gestures. The database consists of annotated gesture regions in the talk transcripts where the speaker performed semantic gestures. Each gesture region is also associated with a video clip of a humanoid avatar performing the semantic gesture. 

The annotation was done by two researchers (authors) who viewed a TED talk video and marked text segments in the transcript where the speaker performed semantic gestures with their arms/hands. A gesture was identified as semantic if it deviated from their natural beat gesture and was representative of the major types of semantic gestures identified in literature (Section ~\ref{sec:gestures_overview}), i.e., iconic, metaphoric, and deictic gestures.  In particular, our annotation aimed to capture the semantic correspondence between a text segment and the associated gesture rather than the text spanning the duration of the gesture. Given the subjective nature of the labeling task, we followed a two-step approach to ensure labeling consistency. Both researchers first calibrated by independently labeling three TED videos and resolving any labeling conflicts through discussions. The final interrater agreement was Cohen’s $\kappa = 0.88$, indicating a strong level of agreement. The remaining videos were annotated by one of the researchers. 

We used the original transcripts uploaded by the TED YouTube channel for annotation. For each TED video, the text transcript was divided into chunks of approximately 100 words, with each chunk forming a sample. Within each sample, text segments where the speaker performed semantic gestures were annotated. A total of 10 TED talk videos (listed in Appendix ~\ref{sec:video_list}) were analyzed, resulting in 252 samples and 994 annotated gesture regions. For each video, we randomly split 80\% of the data for training and 20\% for testing.

\added{The annotated gesture regions have an average length of 3.10 (± 2.48) words with an average of 3.81 (± 2.38) gestures regions in a sample. The gesture video clips in the database were created by enacting the gestures while wearing a Quest Pro headset, with the motion mirrored onto a humanoid avatar using the Movement SDK\footnote{\url{https://developers.meta.com/horizon/documentation/unity/move-overview/}}.  In total, we recorded 343 clips, each 1–3 seconds long, at 1080p resolution and 30 frames per second.}

\subsection{Emphasis Proposal}
\label{model:emph_proposal}
This module takes presenter notes as input and identifies gesture regions that should be emphasized with semantic gestures. For this, we fine-tune an LLM (Llama-2 7B) using a parameter-efficient fine-tuning technique known as Low-Rank Adaptation (LoRA) \cite{hu2022lora}, in which a small number of task-specific trainable parameters (adapter layers) are inserted into the frozen transformer layers. These adapter-layers are trained to learn our specific task of identifying gesture regions. This approach has two-fold benefits. First, using LoRA helps us achieve this task with a few million parameters in a modularized fashion, allowing us to balance the rich knowledge from the pre-trained LLM with the new information learned from the data. Second, this helps reduce the sensitivity towards the system prompt, which can be a challenge in zero-shot prompting, where performance often depends on prompt engineering.

\subsubsection{Fine-tuning with Data Augmentation} Given the limited amount of training data available for fine-tuning, we employed data augmentation using synthetic samples generated by LLMs.  Specifically, we prompted GPT-4o ~\cite{achiam2023gpt} to generate novel text transcripts and their corresponding gesture regions, ensuring they were structurally and semantically similar to the original annotated sample but adapted to different TED talk topics. To maintain data quality, all synthetic samples underwent manual verification to ensure that the generated gesture regions were either semantically equivalent or verbatim to the original annotations. We randomly selected 70\% of the videos for data augmentation, and for each training sample in these videos, we generated five additional augmented samples, increasing the total training dataset size to 881 samples. This augmentation process introduced greater variability into the dataset, enabling the model to better learn the underlying patterns of semantic gestures in text.

\added{We use \texttt{Llama-2 7B}\footnote{\url{https://huggingface.co/meta-llama/Llama-2-7b}} as the base LLM and integrate LoRA adapters with $r=128, \alpha=256$ for fine-tuning using the PEFT\footnote{\url{https://huggingface.co/docs/peft/en/index}} library. The base LLM remains frozen, and only the LoRA parameters are trained. Training is conducted for 20 epochs with a learning rate of $2e-03$ using a single NVIDIA H100 GPU. For fine tuning and subsequent inference, we pass the input text using the standard prompt format\footnote{\url{https://www.llama.com/docs/model-cards-and-prompt-formats/meta-llama-2/}} (seen in Figure ~\ref{fig:model}) of Llama 2 models.}

\subsubsection{Filtering the Predicted Regions.} We include a filtering module that ensures predictions from the \EmphModel{} module align with the original presenter notes before passing them to the \GesModel{} module. This step is important because the generative nature of LLMs can lead to predictions where the gesture regions do not match the original text verbatim. Such differences can range from minor ones (e.g., a missing apostrophe) to complete paraphrasing. The filtering process first attempts a verbatim match between the predicted gesture region and the presenter notes. If no exact match is found, we apply a sliding window approach to search for the most similar gesture region in the notes. Specifically, we compute the cosine similarity between the embedding of the predicted gesture region and the embeddings of each text window in the presenter notes. We set a similarity threshold of 75\%, above which a match is accepted. If no sufficiently similar match is found, the prediction is discarded. We used bge-base-en-v1.5-gguf\footnote{\url{https://huggingface.co/CompendiumLabs/bge-base-en-v1.5-gguf}} as the embedding model.

\subsection{Gesture Identification}
\label{model:ges_iden}
The gesture identification module determines the most appropriate semantic gesture to emphasize the gesture regions predicted by the emphasis proposal model. Selecting the correct semantic gesture for a given gesture region requires understanding the types of gestures available, their contextual meaning, and which gestures are most suitable for emphasis in a particular context.

To systematically model these dependencies, we adopt a Retrieval-Augmented Generation (RAG) framework ~\cite{lewis2020retrieval}, which enables an LLM to access a knowledge base of semantic text–gesture relationships. First, we generate semantic vector embeddings for the labeled gesture regions in the database using a language embedding model. For each gesture region predicted by the emphasis proposal module, we retrieve the most relevant candidates from the vector database based on embedding similarity. We then provide the LLM with the presenter's notes and the retrieved top-k gesture candidates, prompting it to select the most contextually appropriate gesture for emphasizing the given gesture region. The video clip associated with the selected candidate is then used as the gesture cue for that gesture region.

\added{We developed a custom RAG framework using the bge-base-en-v1.5-gguf model to get semantic vector embeddings, FAISS\footnote{\url{https://github.com/facebookresearch/faiss}} to create and efficiently search over the vector database, and the GPT-4o model to generate the final gesture candidate. We retrieve the top-3 gesture candidates using the established approximate K nearest neighbor search with cosine score as the similarity measure ~\cite{kushilevitz1998efficient}. The prompts for LLMs in both modules can be found in Appendix ~\ref{sec:model_details}.}

\section{\System{}: Rehearsal Interface}
\label{system:interface}
Our objective was to identify the essential components and potential benefits of a gesture rehearsal interface. Following Buchenau’s experience prototyping protocol ~\cite{buchenau2000experience}, which emphasizes gaining insights through users' direct interactions with functional prototypes, we developed an initial Wizard-of-Oz protoype of the interface and conducted a pilot study with users.

\begin{figure}[t]%
\centering
\includegraphics[width=0.9\linewidth]{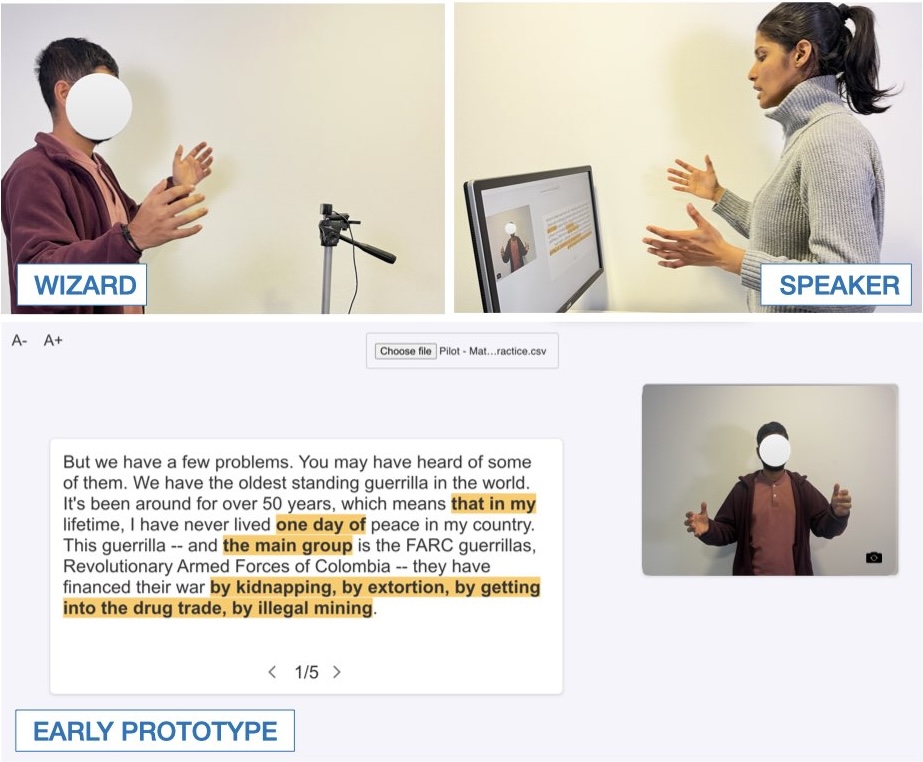}
\caption[]{Early Wizard-of-Oz prototype that simulates a real-time gesture cueing interface for rehearsing talks.}
\Description{This figure visualises the setup for the early wizard-of-oz prototype based pilot study. A wizard is shown performing a gesture in front of a webcam. The speaker sees the demonstrated gesture in near real-time on a early prototype of the rehearsal interface. The speaker mimics the gesture as when the wizard shows a demonstration. A zoomed in view of the early prototype is also shown consisting of the highlighted presenter notes and view of the speaker from the webcam.}
\label{fig:prototype}
\end{figure}

\subsection{Early Prototype}
Our initial prototype was a minimalist web interface developed using ReactJS as shown in Figure \ref{fig:prototype}. It consisted of two windows. The first window displayed the presenter notes that users would rehearse. Text corresponding to the gesture regions were highlighted in the presenter notes. The second window displayed the semantic gesture cue that the user should perform to emphasize the highlighted region. In our prototype, a "wizard" (one of the authors) enacted these cues in real-time via a webcam, displaying them in the gesture cue window.  This setup simulated a just-in-time system that monitors speech and provides guidance to gesture just before users reach the highlighted region. Additionally, the windows were adjustable, allowing users to rearrange and resize them for comfort.  

\subsection{Pilot Study}
We recruited 5 participants (3 males, 2 females, aged 25 to 45 years) who volunteered for the study. \added{Four of them self-identified as novice speakers who had given 1-2 formal talks in the past and one as an expert who had given more than 10 talks.} After an introduction and walkthrough of the prototype, participants were asked to rehearse for a talk, emphasizing the highlighted regions in the presenter notes by mimicking gesture cues displayed in the gesture window. 

We used 5 samples from the database as presenter notes. Participants first read through and understood the text in the presenter notes, before presenting each note with the cued gestures. For rehearsal, they were asked to rehearse each note three times. After the task, participants completed a 15-minute semi-structured interview to discuss the system's usability and share feedback. We also observed participants' interactions throughout the study. Each session lasted approximately 45 minutes.

\subsection{Design Insights \& Final Interface}

All participants recognized the potential of a gesture rehearsal interface as it helped them successfully incorporate the suggested gestures while practicing the talk. Novice speakers, in particular, found it useful for breaking out of a static posture and making their presentations more dynamic: \textit{"It motivates me to do gestures, makes me more relaxed to do gestures" }(P4). Highlighting gesture regions was an effective way for conveying when to gesture. While expert speakers found this highlighting to be sufficient to improve their rehearsal, novice speakers relied on the displayed gesture cues to understand what gestures to use for emphasis. To minimize context switching, they often placed the gesture window next to the notes window. While the interface was useful in its current form, users identified several issues and suggested improvements to enhance the interface. We discuss four key insights, which translate to our final interface.
\begin{itemize}[leftmargin=*, noitemsep, topsep=3pt]
    \item \textbf{Gesture cue visualization.}  \added{Early in our design process, we chose to demonstrate the gestures rather than representing them symbolically. While lightweight symbols such as emojis or gesture icons may be quicker for users to interpret during rehearsal, it can also be challenging to convey the spatial relationships and movement dynamics of gestures with such static representations.} Our pilots indicate that users were generally comfortable following gesture demonstrations presented in a third-person, upper-body view. This perspective allowed them to clearly observe arm and hand movements while understanding their spatial relation to the upper body, aiding effective gesturing. However, two participants reported that guidance from a real human felt distracting, inducing a sense of social evaluation, particularly due to eye contact and facial expressions: \textit{"Its a bit embarrassing ...I feel like someone's looking at me... it make me more self-aware and affects my practice"} (P2). Hence, in our final interface we use a \added{animated} humanoid avatar (upper-body only) captured from below the face, focusing on arm and hand motions to emphasize the gesture.
    
    \item \textbf{Onset of gesture cues.} Effective integration of gestures into speech requires timely presentation of gesture cues. In our pilots, we explored the optimal time for delivering these cues by varying the onset time at which the wizard triggered the cue relative to a highlighted region. We found that presenting gesture cues 4–5 words before the user reached a highlighted region was sufficient for seamless integration.
    
    \item \textbf{Modifying predicted gestures.} Users occasionally found that some of the gesture regions or the suggested gesture did not align with their expectations due to individual communication styles or cultural differences in gesturing norms. Certain gestures may carry different meanings across cultures, making them less suitable or even inappropriate for some users: \textit{"I don't point like this [pointing index finger at audience]... it seems rude... would be good to have other suggestions [for gestures] here"} (P4). To address this, we enable users to customize the assigned gestures by selecting from two alternative gestures retrieved by the gesture identification model.

    \item \textbf{Supporting preview of gestures.} During their initial rehearsal, 3/5 participants indicated difficulty integrating the suggested gestures naturally while speaking. They also struggled with smooth transitions between consecutive gestures. This challenge was particularly pronounced for gestures that participants had not encountered before: \textit{"If I haven't seen the gesture before, I slow down or stop speaking to see the gesture and do it"} (P1). To address this challenge, we introduced a gesture preview feature. By hovering over the highlighted region, users can review and familiarize themselves with the suggested gestures before starting their rehearsal. It also allows users to anticipate transitions between gestures, making their delivery more fluid and natural.
    
\end{itemize}

\subsubsection{Final Interface} Our final interface (Figure ~\ref{fig:teaser}) builds upon our early prototype, incorporating the features identified from our pilots. Users begin by uploading their presenter notes. The gesture recommendation model analyses the notes, predicts gesture regions, and retrieves relevant gestures for each predicted region from the database. These regions are highlighted in the notes and displayed alongside the current presentation slide. Users start their rehearsal by clicking the "Rehearse" button. As they speak, an automatic speech recognition module tracks their progress through the notes. This enables the system to provide real-time gesture cues in the gesture window, offering timely guidance during rehearsal.

The real-time speech tracking module is implemented in Python using the RealtimeSTT library\footnote{\url{https://github.com/KoljaB/RealtimeSTT}} for live speech-to-text conversion. It aligns spoken words with the presenter’s notes using a lexical similarity matching algorithm. Each time the \emph{rehearse} button is pressed, the module resets the speech flow index to the start of the current slide’s notes. During tracking, each newly recognized word prompts a comparison between the last three spoken words and all three-word sequences within a sliding window of the speech notes. The window spans two words before and ten words after the current speech flow index.

Similarity is measured using the Levenshtein similarity ratio \cite{sarkar2016junitmz}, which calculates the minimum number of character edits needed to convert one string to another:
\[
\sigma = \left(1 - \frac{\text{Levenshtein\ Distance}(s_1, s_2)}{\lvert s_1 \rvert + \lvert s_2 \rvert} \right ) \times 100
\]

Here, a 100\% score indicates identical strings. The speech flow index is updated to the position of the phrase with the highest similarity score, provided it meets a minimum threshold of 50\%, which we found to be effective in our pilot study. If no match exceeds this threshold, the index remains unchanged, treating the spoken input as irrelevant to the notes.

\section{Model Evaluation}
We evaluated the performance of our gesture recommendation model in two ways. First, we compared our fine-tuned LLM model with two baseline approaches to evaluate the extent to which our model helps predict gesture regions. We followed this with a more holistic evaluation of our approach through a user study where users subjectively rated the suitability of the predicted gesture regions and retrieved gestures for different samples. 

\subsection{Model Performance}
\subsubsection{Method} We evaluate the \EmphModel{} on the held-out test data from the database (random 80:20 split; 56 samples containing 211 gestures) and on an unseen TED talk video (29 samples containing 256 gestures). We compare our model with two state-of-the-art LLMs: 1) OpenAI GPT-4o model 2) Llama-2 7B. The prompt was kept the same across all models and can be found in the Appendix ~\ref{sec:model_details}.

\subsubsection{Metrics} We compared the recall, precision, and F1 scores across the models using two matching schemes.
\begin{itemize}[leftmargin=*, noitemsep, topsep=3pt]
    \item \textbf{Direct matching:} This is a strict matching where a predicted gesture region is considered a true positive if there is a verbatim overlap with a ground truth.
    \item \textbf{Semantic matching:}  This is a more relaxed matching where predicted gesture regions are considered true positives if there is a 75 percent match in similarity between its vector embedding and that of a ground truth. This is an important metric because the LLM is a generative model and the generated predictions could have minor variations from the original text in terms of spelling, missing letters or punctuation. Gesture regions that are not found in the text are considered invalid and not counted as false positives, as the system directly discards them. 

\end{itemize}

Regardless of the matching scheme, it is essential that the predicted gesture region does not substantially exceed the length of its corresponding ground truth. A model that predicts gesture regions as entire sentences or excessively long phrases may achieve high overlap with the ground truth but lacks the granularity needed for precise gestural guidance. We ensure this constraint by verifying that a predicted region extends no more than three words beyond the matched ground truth.

\begin{table}[t]

\resizebox{\columnwidth}{!}{
\begin{tabular}{@{}lllllllcc@{}}
\toprule
 &
  \multicolumn{2}{c}{\textbf{Precision}} &
  \multicolumn{2}{c}{\textbf{Recall}} &
  \multicolumn{2}{c}{\textbf{F1}} &
  \multicolumn{1}{c}{\multirow{2}{*}{\textbf{\begin{tabular}[c]{@{}c@{}}No. of Predicted \\ gesture regions\end{tabular}}}} &
  \multicolumn{1}{c}{\multirow{2}{*}{\textbf{\begin{tabular}[c]{@{}c@{}}Length of Predictions\\  (No. of words)\end{tabular}}}} \\ \cmidrule(lr){2-7}
 &
  \multicolumn{1}{c}{\textbf{DM}} &
  \multicolumn{1}{c}{\textbf{SM}} &
  \multicolumn{1}{c}{\textbf{DM}} &
  \multicolumn{1}{c}{\textbf{SM}} &
  \multicolumn{1}{c}{\textbf{DM}} &
  \multicolumn{1}{c}{\textbf{SM}} &
  \multicolumn{1}{c}{} &
  \multicolumn{1}{c}{} \\ \midrule
\textbf{GPT-4o}\footnotemark[1]     & 0.034          & 0.039          & 0.043          & 0.047          & 0.038          & 0.043          & 4.18 $\pm {\scriptstyle{1.67}}$ & 15.41 $\pm {\scriptstyle{8.52}}$ \\ \midrule
\textbf{Llama-2 7B} & 0.057          & 0.081          & 0.062          & 0.085          & 0.059          & 0.083          & 4.1 $\pm {\scriptstyle{1.95}}$  & 8.20 $\pm {\scriptstyle{6.52}}$  \\ \midrule
\textbf{Our Model} & \textbf{0.159} & \textbf{0.206} & \textbf{0.185} & \textbf{0.233} & \textbf{0.171} & \textbf{0.219} & 4.45 $\pm {\scriptstyle{1.26}}$ & 2.92 $\pm {\scriptstyle{1.42}}$  \\ \bottomrule
\end{tabular}
}
\caption{Model performance comparison for held-out test samples using direct matching (DM) and semantic matching (SM) schemes.}
\Description{The table shows the precision, recall, and F1 scores for the baselines and our model for held-out samples in the database. Our model has higher scores than the baselines. It also shows average no. of predicted gesture regions in each model and the length of predictions in number of words. Our model predicts similar number of gesture regions (about 4) as off-the-shelf models but is more precise in its predictions with regions being upto 3 words long on average.}
\label{tab:model_results_seen}
\end{table}

\begin{table}[t]

\resizebox{\columnwidth}{!}{
\begin{tabular}{@{}lllllllcc@{}}
\toprule
 &
  \multicolumn{2}{c}{\textbf{Precision}} &
  \multicolumn{2}{c}{\textbf{Recall}} &
  \multicolumn{2}{c}{\textbf{F1}} &
  \multicolumn{1}{c}{\multirow{2}{*}{\textbf{\begin{tabular}[c]{@{}c@{}}No. of Predicted \\ gesture regions\end{tabular}}}} &
  \multicolumn{1}{c}{\multirow{2}{*}{\textbf{\begin{tabular}[c]{@{}c@{}} Length of Predictions\\  (No. of words)\end{tabular}}}} \\ \cmidrule(lr){2-7}
 &
  \multicolumn{1}{c}{\textbf{DM}} &
  \multicolumn{1}{c}{\textbf{SM}} &
  \multicolumn{1}{c}{\textbf{DM}} &
  \multicolumn{1}{c}{\textbf{SM}} &
  \multicolumn{1}{c}{\textbf{DM}} &
  \multicolumn{1}{c}{\textbf{SM}} &
  \multicolumn{1}{c}{} &
  \multicolumn{1}{c}{} \\ \midrule
\textbf{GPT-4o}\footnotemark[1]     & 0.038          & 0.038          & 0.002          & 0.002          & 0.028          & 0.028          & 4.48 $\pm {\scriptstyle{1.13}}$ & 16.40 $\pm {\scriptstyle{9.44}}$ \\ \midrule
\textbf{Llama-2 7B} & 0.190          & 0.230          & 0.101          & 0.114          & 0.132          & 0.152          & 4.17 $\pm {\scriptstyle{2.21}}$  & 6.03 $\pm {\scriptstyle{5.02}}$  \\ \midrule
\textbf{Our Model} & \textbf{0.348} & \textbf{0.490} & \textbf{0.171} & \textbf{0.233} & \textbf{0.230} & \textbf{0.316} & 3.86 $\pm {\scriptstyle{1.25}}$ & 2.93 $\pm {\scriptstyle{1.39}}$  \\ \bottomrule
\end{tabular}
}
\caption{Model performance comparison for unseen TED talk video using direct matching (DM) and semantic matching (SM) schemes.}
\Description{The table shows the precision, recall, and F1 scores for the baselines and our model for an unseen TED video. Our model has higher scores than the baselines. It also shows average no. of predicted gesture regions in each model and the length of predictions in number of words. Our model predicts similar number of gesture regions (about 4) as off-the-shelf models but is more precise in its predictions with regions being upto 3 words long on average.}
\label{tab:model_results_unseen}
\end{table}

\footnotetext[1]{accessed on 21-05-2024}

\begin{figure}[th]%
\centering
\includegraphics[width=0.9\linewidth]{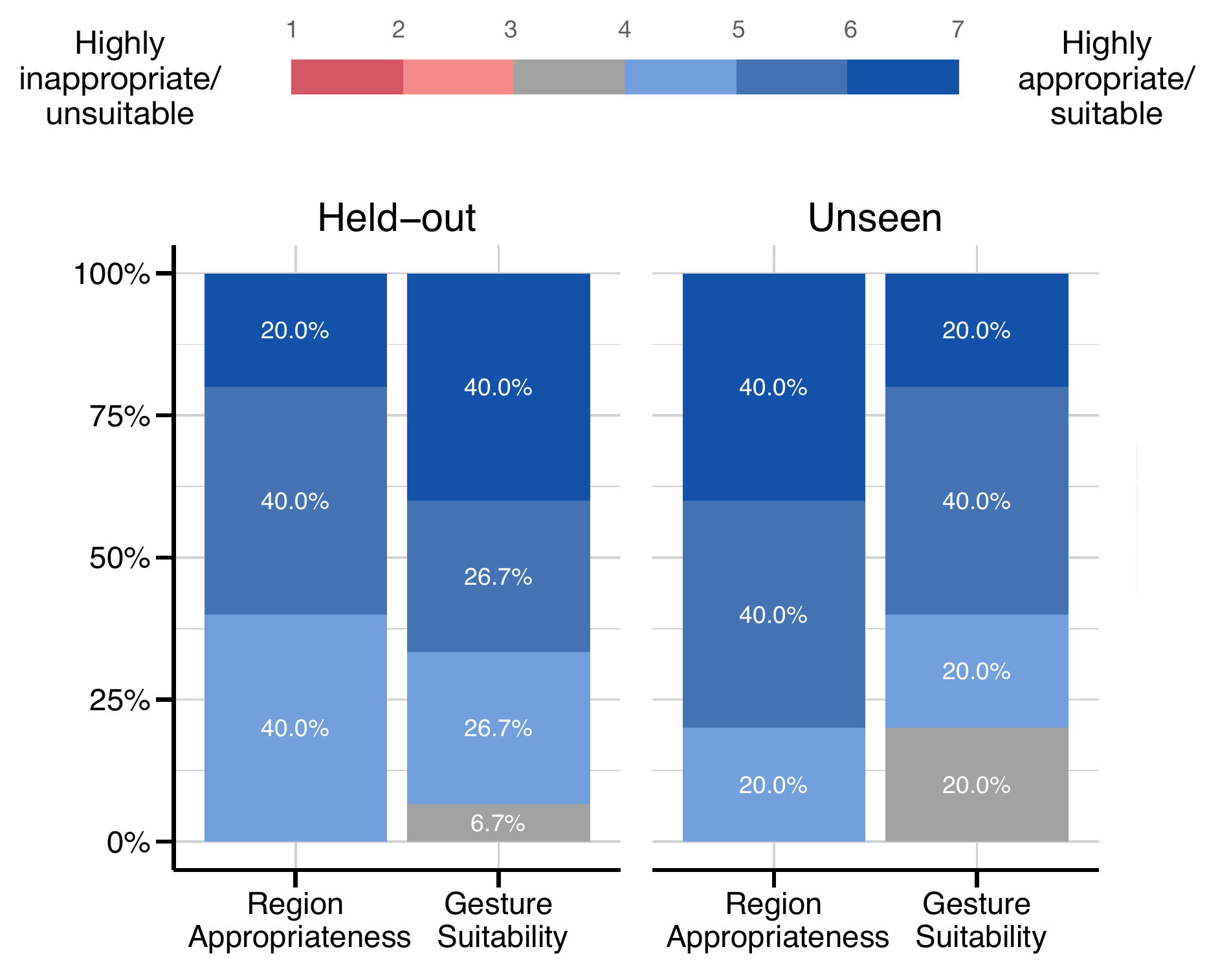}
\caption[]{Distribution of users' ratings for the appropriateness of false positives and the suitability of gestures selected by the model. Over half of false positives were rated as \textit{valid} (avg. ratings >5), and 40\% as \textit{neutral} (avg. ratings 3.5–5). None were clearly inappropriate (avg. ratings < 3), showing the model's recommendations are sensible in general.}
\Description{A stacked 100\% bar graph showing the distribution of ratings for false positive gesture regions. Users ratings for the regions are highly positive along the dimensions of region appropriateness and the semantic gesture suitability, i.e., 60-70\% of the samples have ratings above 5 on a 7 point Likert scale in both dimensions.}
\label{fig:fp_stats}
\end{figure}

\subsubsection{Results} Tables ~\ref{tab:model_results_seen} and ~\ref{tab:model_results_unseen} show that our model outperforms the baseline models across all metrics for both held-out and unseen samples. We see that for held-out samples from our collected dataset (10 videos), our fine-tuned model has an average F1 score of 21.9 in terms of semantic matching, approximately 3 times higher than base Llama-2 model and 6 times higher that GPT-4o. This is mainly because the baseline models lack knowledge of expert gesturing patterns and hence tends to predict overly long gesture regions (upto 16 words long), resulting in significantly lower performance. In contrast, our model learns from expert data and is able to predict more precise regions (upto 3 words long). 

A similar trend is observed on the unseen video, suggesting that our model could generalize well to other TED-style presentations. While the F1 score is higher for this video, a per-video analysis of the mean performance across held-out samples shows that it falls within two standard deviations $(SD = 0.066)$, indicating that the result is consistent with overall model behavior.

Although our model significantly outperforms off-the-shelf LLMs, its performance can appear relatively low when compared to traditional classification tasks. However, we argue that these metrics must be interpreted in the context of the task. Even though there are linguistic elements in spoken discourse where, as shown in previous research and also observed in our dataset, expert speakers commonly tend to gesture — for instance, when referring to themselves or the audience, during topic shifts such as discourse connectives, or when listing items — there are also notable speaker variations. Some speakers, for example, tend to perform iconic gestures such as mimicking scissors for the word "cut." These gestures form the long tail of semantic gestures and account for a very small percentage of the dataset. Due to this imbalanced distribution, the model tends to learn the more frequently occurring gestures. This is a common challenge in machine learning models that deal with long tail distributions. While our model effectively captures broad gesturing patterns shared across speakers, it may fail to predict less frequent or speaker-specific gestures, leading to lower recall.

Similarly, the low precision scores do not necessarily indicate poor performance. Many of the false positives could still be valid gesture regions, suggested based on patterns learned from other speakers. To better understand this, we conduct a user study to subjectively evaluate the model predictions, particularly false positives, in the following section.

\subsection{Quality of Predicted Gestures}

While our model outperforms off-the-shelf LLMs in predicting gesture regions, the low precision scores raise an important question: Are the detected false positive regions truly inappropriate for gestures, or are some still valid for gesturing? It is possible that a predicted region was not gestured by the original speaker but could still serve as a meaningful emphasis point if accompanied by a suitable semantic gesture. To explore this, we conducted a user study to subjectively evaluate the quality of the gesture regions predicted by our \EmphModel{} model and the suitability of the gestures selected by our \GesModel{} model.

\subsubsection{Method} The study involved viewing video clips of a person who enacted a presentation of the sample text, emphasizing regions predicted by our \EmphModel{} model using the gestures retrieved by the \GesModel{} model. Participants watched the video clips and evaluated each gesture region on a 7-point Likert scale based on two criteria: (a) the appropriateness of emphasizing the region during the presentation with a gesture, and (b) the suitability of the gesture used to emphasize that region. 

We randomly selected five samples (containing 18 gestures) each from the held-out test data  in the gesture database and from the unseen video (containing 16 gestures). The video clips were enacted by one of the authors with their face blurred to ensure anonymity and prevent any potential bias in ratings. We recruited 30 (16 female, 13 male, 1 other) volunteers with a mean age of 28.2 (SD=5.19) years from the university and randomly assigned them to evaluate either the held-out or the unseen video set. Evaluating a video set took approximately 15 minutes.

\subsubsection{Results} Overall, participants found the gesture regions proposed by the \EmphModel{} model appropriate for emphasis in both held-out ($M=5.23, SD=1.59$) and unseen samples ($M=5.56, SD=1.43$). Similarly, they rated the semantic gestures retrieved by the \GesModel{} model as well-suited for emphasis in both held-out ($M=5.44, SD=1.58$) and unseen samples ($M=5.08, SD=1.55$). To assess the impact of false positives (see Figure ~\ref{fig:fp_stats}), we define “valid” false positives as samples with an average rating of higher than 5 in both appropriateness and suitability. These can be considered true positives, as users found them to be sensible emphasis points. We found that 53.3\% of false positives in held-out samples and 60\% in unseen samples met this criterion. Additionally, 40\% of false positives in both cases scored between 3.5 and 5. Such samples can be considered to be “neutral” or slightly positive, implying gesturing at those points would not negatively affect the presentation. Notably, no samples scored below 3, indicating that the model does not suggest completely irrelevant or unsuitable gesture regions. \added{These results highlight the importance of subjective evaluation in generative tasks like gesture prediction, where traditional metrics such as F1 score or segmentation accuracy may not fully capture the quality or effectiveness of the model predictions.}

\section{User Evaluation}
We conducted a user study with \numusers{} participants to evaluate whether rehearsing presentations with \System{} can help users positively affect their gesturing while giving the talk. We also evaluate the user experience and usability of the system to understand its strengths and limitations.

\subsection{Method}
We conducted a within-subject study with two counterbalanced conditions: (1) a baseline \textit{Notes} condition, where participants rehearsed using only the presenter notes, and (2) the \System{} condition, where participants viewed recommended gesture regions and received proactive gesture cues. The rehearsal interface for Notes condition was identical to the \System{} condition, but without the highlighted regions and gesture cues. To minimize learning effects, we used two separate parts of the transcript from the unseen TED talk video for each condition. We segmented the transcript into 100-word chunks to create the presenter notes for each slide. The slide sets were counterbalanced across the conditions. Participants rehearsed five slides per condition, rehearsing each slide three times. After rehearsal, participants presented the talk as they would in front of an audience, using just the presenter notes without any intervention.

After obtaining informed consent, we briefed participants on the task and asked them to rehearse and deliver two talks as if presenting to an audience. In the \System{} condition, participants were first introduced to the system and its functionalities. During rehearsal, they could modify or delete the recommended gestures based on their preference. Their presentations after rehearsal were video-recorded for analysis. After each condition, participants reviewed the recorded presentation of their talk. The study concluded with a short survey comparing the two rehearsal methods followed by semi-structured interview to gather qualitative feedback.

\subsection{Participants}
Our participants included 4 males and 6 females, who were proficient in English with an average
age of 25.42 years. Their expertise in giving talks in front of an audience varied; 1 participant identified as beginner (1-2 talks), 2 as intermediate (3-5 talks), 5 as advanced (5-10 talks), and 2 as experts (10+ talks). Each participant spent 1 hour
in the study and received 10 Euros for their time.
\begin{figure}[th]%
\centering
\includegraphics[width=\linewidth]{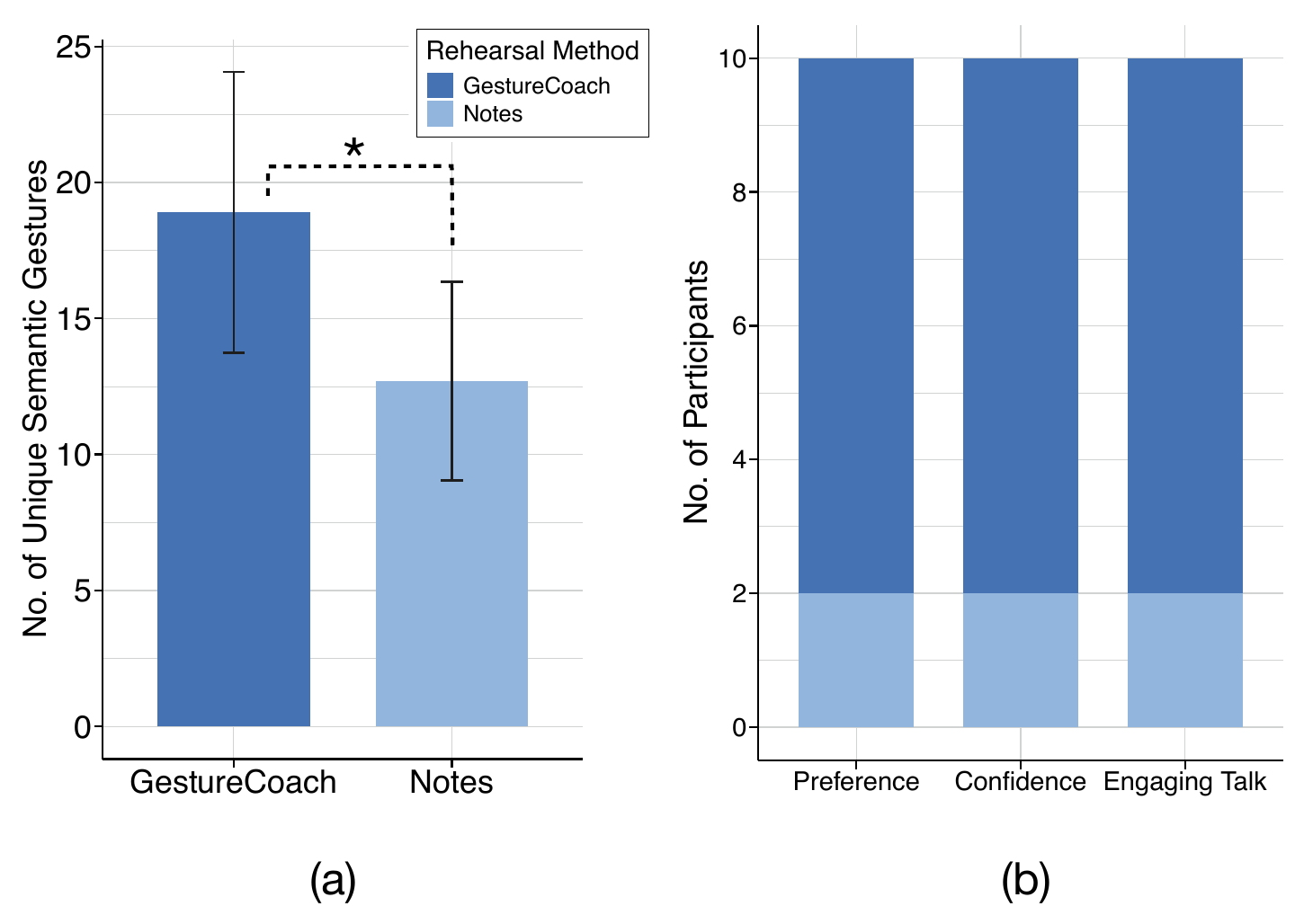}
\caption[]{Comparison of gesture usage and user preference after rehearsing with \System{} vs. Notes.  \System{} led to significantly more unique semantic gesture use and was preferred by 8 of \numusers{} participants, who reported increased confidence after practice and a more engaging final talk. }
\Description{Figure (a) shows a bar graph comparing the average number of unique semantic gestures performed after rehearsing with GestureCoach vs. Notes. GestureCoach is significantly higher than Notes. Figure (b) shows a stacked bar graph of preferences between the two rehearsal methods. 8/10 users preferred GestureCoach overall and found that it improved their confidence and helped deliver a more engaging talk as compared to rehearsing with just Notes.}
\label{fig:results_1}
\end{figure}

\begin{figure*}[th]%
\centering
\includegraphics[width=\linewidth]{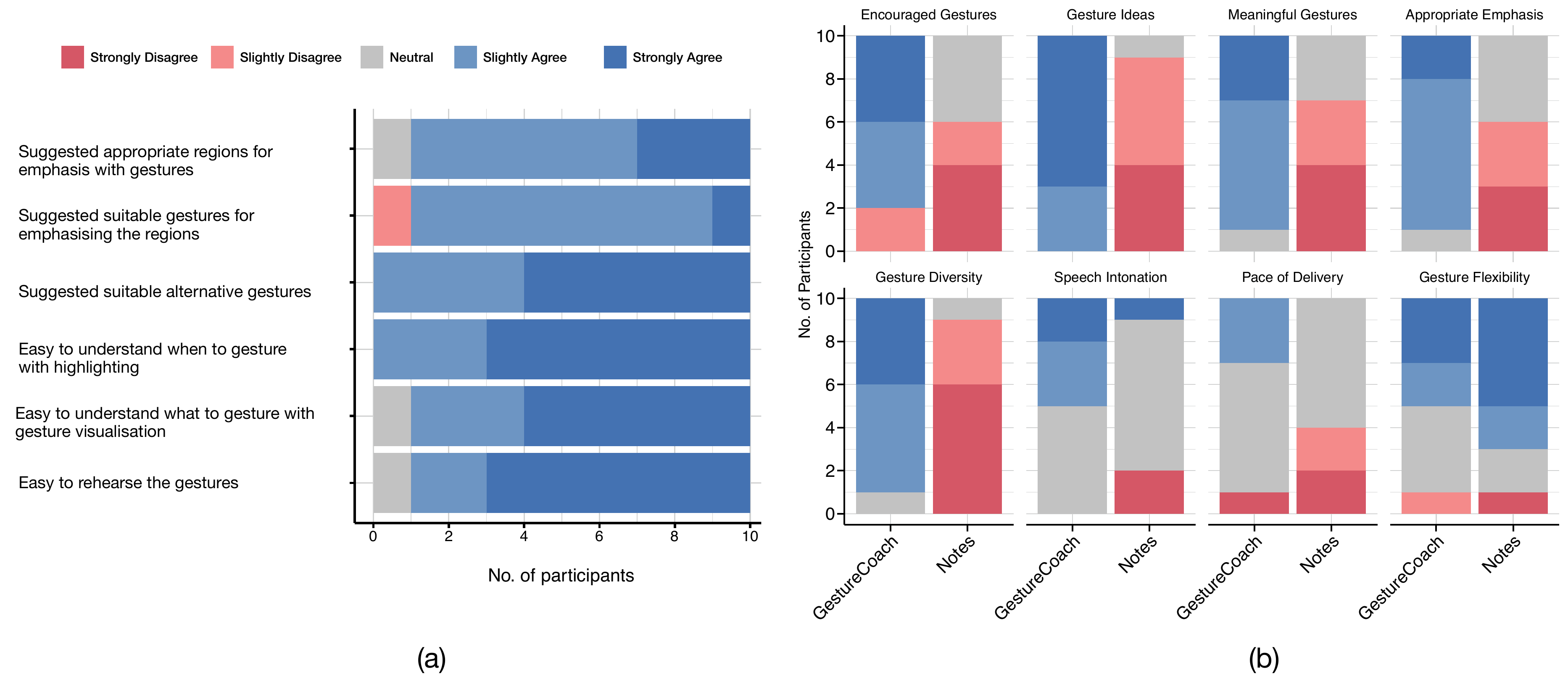}
\caption[]{(a) Subjective evaluation of \System{} features indicate its usefulness and ease of use for rehearsal, (b) Results of subjective comparison between \System{} and Notes.}
\Description{Figure (a) shows a horizontal bar graph of the rating distribution on a 5 point Likert scale for various features of the GestureCoach System. Most users agree strongly that the features helped them during rehearsal. Figure (b) a vertical bar graph comparing the rating distribution for GestureCoach and Notes for different aspects such as encouraging gestures, providing gesture ideas, speech intonation etc. Most users strongly agree that GestureCoach improved these aspects while Notes did not. Please refer to the text for specific details.}
\label{fig:results_2}
\end{figure*}

\subsection{Results}

\subsubsection{Gesture Quantity} We quantified semantic gestures performed by speakers during their talks. An expert in the gesture domain (co-author) blind to the conditions annotated the recorded talks using the same criteria used for creating the gesture database (Section ~\ref{sec:database}). Overall, speakers performed slightly higher number of gestures in the \System{} condition $(M=30.30, SD=8.87)$ than in the Notes  $(M=25.1, SD=10.16)$ condition, but the difference was not significant. A closer look at speakers' individual data showed that 8 of \numusers{} speakers performed significantly more gestures $(t(7) = 2.99, p = .02)$ after rehearsing with \System{} $(M=31.75, SD=8.97)$ than with Notes $(M=21.65, SD=7.52)$. The remaining two users showed an opposite trend, performing more gestures after using Notes $(M=39, SD=7.07)$ than with \System{} $(M=24.5, SD=7.77)$. We discuss potential reasons for this in Section ~\ref{sec:impact_on_natural_gestures} based on users' qualitative feedback.

\subsubsection{Gesture Diversity} We measure gesture diversity as the number of unique semantic gestures performed by speakers during their talk. As seen in Figure ~\ref{fig:results_1}a, speakers performed considerably higher number of unique gestures in the \System{} condition $(M=18.90, SD=7.21)$ than in the Notes  $(M=12.70, SD=5.10)$ condition. A paired samples t-test showed that the difference is significant $(t(9) = 3.27, p = .009)$ with a large effect size, Cohen’s $d = 1.03$, suggesting that rehearsing with \System{} helped speakers add diversity to the gestures they used during their talk. 

\subsubsection{Subjective Evaluation of \System{}} Overall, speakers found the \System{} interface intuitive and easy to use with an average SUS score of 90.83 $(SD=5.86)$. Figure ~\ref{fig:results_2}a shows speakers' evaluation of \System{} using a 5-point Likert scale. They found that the system recommended appropriate gesture regions for emphasis $(M=4.20, SD=0.63)$ and that the suggested gestures aligned well with the emphasized text $(M=3.9, SD=0.73)$. In cases where the original gesture was unsuitable, speakers found that the alternatives recommended by \System{} often included a suitable one $(M=4.6, SD=0.51)$.  Using highlights $(M=4.70, SD=0.48)$ and gesture visualizations $(M=4.50, SD=0.70)$ to represent the suggestions were found to be easy to understand for users. There was also strong agreement among speakers that rehearsal experiences with these gestures was easy $(M=4.6, SD=0.69)$.

\subsubsection{Subjective Comparison of Rehearsal Methods} Rehearsal with \System{} was the preferred method for most speakers (8/\numusers{}), enhancing their confidence (8/\numusers{}) and enabling them to deliver a more engaging talk (8/\numusers{}) compared to rehearsing with Notes. As shown in Figure ~\ref{fig:results_2}b, speakers reported that using \System{} during rehearsal encouraged them to incorporate gestures more naturally $(M=4, SD=1.15)$ \added{$(t(9) = 3.87, p < .01)$}, provided ideas for gesture use $(M=4.70, SD=0.48)$ \added{$(W = 55.0, z = 2.80, p < .01)$}, and helped them use meaningful gestures $(M=4.20, SD=0.63)$ \added{$(W = 55.0, z = 2.80, p < .01)$} at appropriate points $(M=4.1, SD=0.56)$ \added{$(t(9) = 6.00, p < .001)$}. It also added diversity to the gestures $(M=4.30, SD=0.67)$ \added{$(t(9) = 7.79, p < .001)$} during their talk. Additionally, some participants observed improvements in their intonation $(M=3.70, SD=0.82)$ \added{$(t(9) = 1.96, p = 0.081)$} and pacing $(M=3.10, SD=0.87)$ \added{$(W = 6.0, z = 1.60, p = 0.174)$} after rehearsing with \System{}. However, they noted that rehearsing with Notes offered more flexibility $(M=4, SD=1.33)$ for integrating their own gestures than \System{} $(M=3.70, SD=1.06)$, \added{although the difference was not significant}.
\section{Discussion}
In the following section, we discuss the main insights from the user evaluation of \System{} and derived design implications.

\subsection{Summary of Findings}

\subsubsection{Rehearsing with \System{} Helps
Deliver More Engaging Talks} Overall, participants reported that rehearsing with \System{} enhanced their ability to deliver a more dynamic and expressive presentation. Speakers perceived an increase in gesture usage and the suitability of gestures used after rehearsing with \System{}. This perception is also supported by our analysis of speakers' gesture usage, which showed a 48.8\% increase in the diversity of gestures used when speakers rehearsed with \System{} compared to using Notes.

Feedback from speakers indicates that the gesture recommendations provided by \System{} during rehearsal played a key role in encouraging them to do more gestures during their talk. As P1 described, focusing on gestures when rehearsing with just notes was challenging: \textit{“[Without \System{}] I know I need to move [my hands], but it’s difficult to think about what would be a good gesture, so I end up just doing random gestures"}. In contrast, P1 explained that \System{} \textit{"helped me reflect on where to emphasize and what kind of gestures I could use... it's like someone giving feedback"}. The recommendations provided practical examples and demonstrated how to use gestures more intentionally.

While the primary objective of \System{} was to enhance users' gesturing during presentations, some speakers also reported a synergistic effect on their speech delivery. For instance, P7 felt that rehearsing with \System{} improved their intonation: \textit{"When it's highlighted I also stress that point in my speech such that it matches the gesture"}. This positive impact on intonation occurs because effectively performing a gesture with the intended emphasis requires speakers to slow down and emphasize the associated words, allowing the gesture’s intended meaning to be conveyed more effectively. Thus, by encouraging intentional emphasis in both gestures and speech, \System{} helped speakers deliver more expressive and engaging talks.

\subsubsection{Enhancing Speakers' Gesture Vocabulary}

A key goal of \System{} was to enhance speakers’ gesture repertoire by providing access to the rich gesture vocabulary used by experts. Participant feedback strongly indicates that \System{} successfully achieved this goal by encouraging speakers to go beyond their usual, simplistic gestures and explore a wider range of expressive movements. The ability to quickly preview gestures and explore alternative ones suggested by \System{} was particularly valuable in this regard. We observed that speakers frequently browsed through these gesture recommendations, even when the initially recommended gesture was already meaningful: \textit{“I’m curious on what other gestures I could use... It gives me options”} (P1). This exploration evolved into an active reflection process, where speakers considered the suitability of each gesture and selected one that aligned with their expectations: \textit{"It made me think more on what gesture I should use.. I consider the options and select the best"} (P10). Through this process, speakers were exposed to new gestures, adding diversity to their repertoire: \textit{“I normally do a lot of simplistic gestures, but it [\System{}] gave me really iconic or unique gestures that fit the context and add value to the talk”} (P5). 

\subsubsection{Learnability of Gestures} Speakers found it easy to incorporate the gestures within their talk during rehearsal and later recall many of them without any highlighted regions or gesture cues. P2 attributed this to the strong semantic relevance between the gestures and the emphasized text: \textit{“The suggested gestures contextually make sense, so it was easy to learn and do it later”}. This meaningful connection allowed speakers to naturally associate the gestures with the corresponding text in their presenter notes, making the gestures easier to remember and perform. These findings suggest that targeted semantic gesture recommendations could have significant potential to improve gesture training compared to generic feedback about the speaker's amount of gesturing \cite{schneider2015presentation, damian2015social, barmaki2016kinesic}. 

\subsubsection{Structured Rehearsal Improved Confidence}

Traditional rehearsal often left speakers uncertain about what gestures to incorporate, relying primarily on gestures they could recall from their previous talks. As P4 explained, \textit{"The gestures I use are too random when I rehearse on my own. It's different in each iteration"}. This lack of structure was challenging for moderately experienced speakers and even more so for those who are naturally less expressive. In contrast, \System{} enabled speakers to critically reflect on the gestures they intended to use during their presentations. This reflection process allowed them to create an initial mental "sketch" of their talk, which structured their rehearsal process: \textit{"gave me a consistent set of gestures that I can start with"} (P4). This structured rehearsal made it easier to rehearse with \System{} and ultimately enhanced participants’ confidence in delivering their talks post-rehearsal.

\subsection{Design Implications}

\subsubsection{Personalization via Human-AI Collaborative Authoring}

Participants' ratings of the gesture recommendations provided by \System{} indicate that they generally found the gesture regions to be appropriate and effectively paired with meaningful gestures. However, participants did not merely accept the originally suggested gestures passively. Instead, the process of identifying the best gesture to use for a recommended region was iterative and involved active reflection and modification. Through this iterative process, participants engaged in a collaborative authoring experience with \System{}, combining inputs from the system with their own experiences to create a gesture pallette that felt unique and personalized to their presentation style. These findings align with insights from evaluations of recent AI-driven interfaces \cite{qian2024shape, su2024sonify}, which emphasize the importance of providing customization controls and ensuring that the human remains actively involved in the creative process. Future research could further enhance this collaboration by enabling users to refine the system's recommendations through prompts or demonstrations, and support gesture palette creation by actively learning from user selections.

\subsubsection{Hybrid Rehearsal to Integrate with Natural Gestures}
\label{sec:impact_on_natural_gestures}
While \System{} helped speakers learn and incorporate diverse gestures into their presentations, expert speakers reported an unintended side effect: training with the system appeared to suppress their natural "beat" gestures, particularly during rehearsal. For these speakers, the recommendations and structured guidance provided by \System{} proved to be a double-edged sword. On the one hand, the system accelerated the rehearsal process and offered valuable inspiration, as noted by P4: \textit{“It speeds up the process my rehearsal and gives me these ideas... But, it also constrains me because I focus too much on the suggested regions and getting them right that I don't focus much on other gesturing at parts of my speech”}. To address this, these speakers expressed that they would use \System{} in a hybrid manner during rehearsal: beginning with the system to explore gesture suggestions,  then continuing to rehearse using only the highlighted regions. This approach would allow them to integrate \System{}’s suggestions while adapting them to their own natural gesturing style. Future rehearsal systems could better support this process by observing speakers' gesturing and providing cues only at targeted points in a unobtrusive manner. In addition, participants expressed a preference for being able to highlight new text segments they felt should be emphasized and to receive corresponding gesture recommendations from the system.

\subsubsection{\added{Modeling Considerations for Gesture Recommendation}}
\added{We approached the gesture recommendation problem as two distinct tasks: deciding \emph{when} to gesture and determining \emph{what} gesture to use. The \emph{when} component was driven by observable and well-established patterns such as gestures frequently occurring at topic changes, during list enumerations, or at points of emphasis. These patterns make it easier to model it as a finetuning task. In contrast, the \emph{what component} was more complex due to high variability. Semantic gestures follow a long-tail distribution, where a single gesture can suit many different spoken contexts. For such problems, a RAG-based approach may be a better choice as it enables the model to retrieve grounded examples during inference rather than relying entirely on mappings learned through direct fine-tuning. 
}

\subsection{Limitations \& Future work }

\subsubsection{Need for verbatim presenter notes} Our gesture recommendation model requires users to provide a complete transcript of their talk in order predict gesture regions and potential gestures. During our pilots with the early prototype, novice speakers indicated that they indeed prefer to create such detailed presenter notes and closely adhere to them, particularly for timed presentations. In contrast, expert speakers typically rely on brief bullet points as a reference rather a full script. Future research should explore ways to relax the requirement for presenter notes while still capturing enough information to make accurate gesture predictions.

\subsubsection{Guidance for subtleties in gestures} Our system currently provides only a general representation of a gesture’s overall motion but does not capture finer details such as the speed or intensity with which the gesture should be performed, which may depend on the context of the talk. Additionally, the system does not account for other aspects of body language, such as head movements or body posture, which are often used to enhance talks. Future gesture recommendation systems should aim to model these nuances and incorporate them into gesture cues to provide more comprehensive guidance during rehearsal.

\subsubsection{\added{Design Space of Gesture Visualizations}} \added{In our system, we visualized gesture cues using an animated avatar that demonstrated the gesture motion. However, it is also possible to consider other lightweight alternatives, such as emojis or icons, that are embedded within the presentation script. Such cues maybe easier to reference while speaking and could be particularly helpful in real-time contexts. Future work should explore the design space of gesture visualizations and empirically compare them to understand the trade-offs between gesture comprehension and cognitive load.}

\subsubsection{\added{Adapting to cultural contexts}} The gesture recommendation model selects gestures from a database of semantic gestures collected from expert speakers of diverse cultural backgrounds. However, our studies revealed that some gestures may not be appropriate for users from specific backgrounds. To address this, we provide customization options that allow users to choose alternative gestures for a given region. An improved system could further enhance this adaptability by actively learning user preferences through feedback, enabling more personalized gesture recommendations over time.

\subsubsection{Generalisability beyond TED-style presentations} Our \EmphModel{} module is fine-tuned on data from TED talk videos and shows promising predictive performance on both held-out and unseen presenter notes of TED-style talks. Although we are optimistic that it could generalize to other settings, such as public speeches, differences in gesturing patterns across contexts may not be fully captured. Future work should focus on developing more generalizable models by training on larger, more diverse datasets or creating models that adapt gesture predictions depending on the speaking context.

\section{Conclusion}

In this paper, we introduced \System{}, a system designed to proactively guide speakers during rehearsal in using effective, well-timed semantic gestures in their talk.  To realise this, we developed an LLM-driven gesture recommendation model that predicts appropriate gesture regions from presenter notes and suggests suitable semantic gestures based on expert speakers’ gesturing patterns. Our model evaluation shows promising accuracy in predicting gesture regions and suitable gestures. Building on insights from an exploratory study using a Wizard-of-Oz prototype, we then designed a rehearsal interface that proactively cues speakers to integrate the gesture recommendations from the model during rehearsal. An evaluation of our implemented system with users highlighted its efficacy in encouraging speakers to perform more diverse gestures, leading to a more engaging talk. These findings open up new possibilities for AI-driven rehearsal tools to support speaker training and performance.

\begin{acks}
This work has been partly supported by the Deutsche Forschungsgemeinschaft DFG under Grant No. 521601028 within the Priority Program SPP 2199 Scalable Interaction Paradigms for Pervasive Computing Environments, and the Deutsche Forschungsgemeinschaft, Funder Id: \url{http://dx.doi.org/10.13039/501100001659}, SFB 1102: ``Information Density and Linguistic Encoding'', project number 232722074. We also thank the reviewers for their insightful comments that helped to improve the paper.
\end{acks}

\bibliographystyle{ACM-Reference-Format}
\bibliography{sample-base}

\appendix

\section{List of TED videos}
\label{sec:video_list}

\begin{table}[ht]
\centering
\resizebox{\columnwidth}{!}{
\begin{tabular}{p{9cm}p{6cm}}
\hline
\textbf{TED talk Title} & \textbf{URL} \\
\hline
Jose Miguel Sokoloff: How we used Christmas lights to fight a war | TED & \href{https://www.youtube.com/watch?v=0Fi83BHQsMA}{\texttt{youtube.com/watch?v=0Fi83BHQsMA}} \\
\hline
How to stay calm when you know you'll be stressed | Daniel Levitin | TED & \href{https://www.youtube.com/watch?v=8jPQjjsBbIc}{\texttt{youtube.com/watch?v=8jPQjjsBbIc}} \\
\hline
Why good leaders make you feel safe | Simon Sinek | TED & \href{https://www.youtube.com/watch?v=lmyZMtPVodo}{\texttt{youtube.com/watch?v=lmyZMtPVodo}} \\
\hline
This could be why you're depressed or anxious | Johann Hari | TED & \href{https://www.youtube.com/watch?v=MB5IX-np5fE}{\texttt{youtube.com/watch?v=MB5IX-np5fE}} \\
\hline
What you need to know about CRISPR | Ellen Jorgensen & \href{https://www.youtube.com/watch?v=1BXYSGepx7Q}{\texttt{youtube.com/watch?v=1BXYSGepx7Q}} \\
\hline
Why I love a country that once betrayed me | George Takei & \href{https://www.youtube.com/watch?v=LeBKBFAPwNc}{\texttt{youtube.com/watch?v=LeBKBFAPwNc}} \\
\hline
Wendy Suzuki: The brain-changing benefits of exercise & \href{https://www.youtube.com/watch?v=BHY0FxzoKZE}{\texttt{youtube.com/watch?v=BHY0FxzoKZE}} \\
\hline
What You Can Do to Prevent Alzheimer's | Lisa Genova & \href{https://www.youtube.com/watch?v=twG4mr6Jov0}{\texttt{youtube.com/watch?v=twG4mr6Jov0}} \\
\hline
How to build your confidence -- and spark it in others | Brittany Packnett Cunningham & \href{https://www.youtube.com/watch?v=b5ZESpOAolU}{\texttt{youtube.com/watch?v=b5ZESpOAolU}} \\
\hline
Why you should define your fears instead of your goals | Tim Ferriss & \href{https://www.youtube.com/watch?v=5J6jAC6XxAI}{\texttt{youtube.com/watch?v=5J6jAC6XxAI}} \\
\hline
\\
How craving attention makes you less creative | Joseph Gordon-Levitt  \footnotemark[1] & \href{https://www.youtube.com/watch?v=3VTsIju1dLI}{\texttt{youtube.com/watch?v=3VTsIju1dLI}} \\
\hline
\end{tabular}
}
\label{tab:ted-talks}
\end{table}

\section{Model details}
\label{sec:model_details}

\footnotetext[1]{Unseen video used only for testing}

\subsection{System Prompt for \emph{Emphasis Proposal} module}
\label{sec:model_emph_prompt}

\begin{lstlisting}
You are an expert in public speaking. The provided Text will be used for a TED talk. Identify the Emphasis Areas in the Text that should emphasized when giving the TED talk.The extracted text in the Emphasis Areas should exactly match phrases in the given Text.
\end{lstlisting}
\added{\textbf{Example Prompt:} <s>[INST] <<SYS>> You are an expert in public speaking. The provided Text will be used for a TED talk. Identify the Emphasis Areas in the Text that should emphasized when giving the TED talk.The extracted text in the Emphasis Areas should exactly match phrases in the given Text. <</SYS>> Text: Cortisol is toxic, and it causes cloudy thinking. So part of the practice of the pre-mortem is to recognize that under stress you're not going to be at your best, and you should put systems in place. And there's perhaps no more stressful a situation than when you're confronted with a medical decision to make. And at some point, all of us are going to be in that position, where we have to make a very important decision about the future of our medical care or that of a loved one, to help them with a decision. And so I want to talk about that. [/INST]}

\subsection{System Prompt for \emph{Gesture Identification} module}
\label{sec:model_ges_prompt}

\begin{lstlisting}
You are an expert in public speaking. You are given a TEXT and a QUERY phrase within the text that needs to be emphasized. Additionally, you are provided with CANDIDATE emphasis areas" Your task is to identify the CANDIDATE emphasis phrase that is most similar to the QUERY phrase in the TEXT in terms of emphasis context. Always choose only from the provided list of CANDIDATES.
\end{lstlisting}

\added{\textbf{Example Prompt:} You are an expert in public speaking. You are given a TEXT and a QUERY phrase within the text that needs to be emphasized. Additionally, you are provided with CANDIDATE emphasis areas" Your task is to identify the CANDIDATE emphasis phrase that is most similar to the QUERY phrase in the TEXT in terms of emphasis context. Always choose only from the provided list of CANDIDATES. TEXT: First of all, thank you for your attention. There's nothing quite like being in a room full of people like this, where all of you are giving your attention to me. It's a powerful feeling, to get attention. I'm an actor, so I'm a bit of an expert on, well, nothing, really. But I do know what it feels like to get attention -- I've been lucky in my life to get a lot more than my fair share of attention.; QUERY: nothing; CANDIDATES: 1) there's nothing 2) no matter, 3) the only thing. Give only the candidate as output.'}

\end{document}